\documentclass[pre,twocolumn,aps,eqsecnum]{revtex4}
\usepackage{epsfig}
\pdfoutput=1
\begin{document}

\title{Theories of Glass Formation and the Glass Transition}

\author{J.S. Langer}
\affiliation{Department of Physics, University of California, Santa Barbara, CA  93106-9530}

\date{\today}

\begin{abstract}
This key-issues review is a plea for a new focus on simpler and more realistic models of glass-forming fluids.  It seems to me that we have too often been led astray by sophisticated mathematical models that beautifully capture some of the most intriguing features of glassy behavior, but are too unrealistic to provide bases for predictive theories.  As illustrations of what I mean, the first part of this article is devoted to brief summaries of imaginative, sensible, but disparate and often contradictory ideas for solving glass problems.  Almost all of these ideas remain alive today, with their own enthusiastic advocates.  I then describe numerical simulations, mostly by H. Tanaka and coworkers, in which it appears that very simple, polydisperse systems of hard disks and spheres develop long range, Ising-like, bond-orientational order as they approach glass transitions.  Finally, I summarize my recent proposal that topologically ordered clusters of particles, in disordered environments, tend to become aligned with each other as if they were two-state systems, and thus produce the observed Ising-like behavior.  Neither Tanaka's results nor my proposed interpretation of them fit comfortably within any of the currently popular glass theories.     

\end{abstract}

\maketitle

\section{Introduction}
\label{Intro}

In 2007, when I published a Reference Frame column in {\it Physics Today} about ``the mysterious glass transition,'' \cite{JSL-RefFrame-07} I was just beginning to realize how deeply divided the field of glass physics had become. Not much has changed since then, despite the emergence of interesting new information.  This article contains an extension of the opinions I expressed in that essay, plus a summary of the new information and a proposed strategy for moving forward. 

It is remarkable that, after many decades of intense study, there is still no generally accepted, fundamental understanding of glassy states of matter or the processes by which they are formed.  We know that a wide variety of liquids -- molecular, metallic, colloidal, etc. -- can be cooled into stable or very long-lived metastable states in which they remain noncrystalline.  As such a system becomes colder, its viscosity increases dramatically, as if its internal relaxation mechanisms were controlled by thermally  activated processes whose barrier heights grow with decreasing temperature. When observed in more detail, it is seen to undergo anomalously slow, ``stretched exponential'' relaxation in its approach to equilibrium.   Ultimately, such a material falls out of thermodynamic equilibrium and becomes ``glassy'' below a temperature $T_g$, defined by Angell and others \cite{ANGELL-95} in terms of an arbitrarily chosen, very long, time scale.  This material may, or may not, undergo a sharp dynamic glass transition at an experimentally inaccessible temperature $T_0 < T_g$, where the viscosity extrapolates to infinity.  Glassy states at temperatures below $T_g$ generally exhibit various kinds of aging processes; but they are solidlike in the sense that they seem to be infinitely viscous and to possess nonzero shear moduli.  

In addition to this dynamic signature, glass transitions exhibit characteristic thermodynamic properties. Kauzmann showed in 1948 \cite{KAUZMANN-48} that the entropy deduced from a specific heat measurement in a high-temperature glass-forming material appears to extrapolate down to a value comparable to the entropy of the corresponding crystalline state at a thermodynamic transition temperature $T_K$, roughly the same as the dynamic temperature $T_0$.  The specific heat jumps irreversibly from lower to higher values with increasing temperature near the transition point.  Thus, some appreciable fraction of the degrees of freedom of a glassy material appears to be frozen.  The glassy state is non-ergodic; it somehow is unable to explore a statistically significant fraction of its configuration space on experimental time scales. Upon reheating, the frozen degrees of freedom are reactivated and the specific heat rises abruptly. 

These basic features of glass transitions have been tested in a wide range of laboratory observations and numerical simulations.  They have been brought into sharper focus by measurements of frequency dependent viscoelastic and dielectric responses, self intermediate scattering functions, vibrational spectra, and the like; and they have been observed in direct images of particle motions in colloidal suspensions and numerical simulations.  

However, there is one overwhelmingly important difficulty that faces all of these observations, specifically, the dramatic increase in equilibration times that occurs near glass transitions.  Glassy slowing down is qualitatively more extreme than its analog in fluid or magnetic critical phenomena; thus it has been extraordinarily difficult to probe the most fundamental aspects of equilibrium glass physics by either experimental or computational techniques.  This natural obstacle to progress is a major reason why the community of glass scientists has had so much trouble understanding one of nature's most important classes of materials phenomena.  

Nevertheless, some developments starting in 2010 -- especially numerical simulations by Tanaka and coworkers \cite{TANAKAetal-10,KAWASAKI-TANAKA-11,TANAKA-EPJ-12}, independently corroborated by others \cite{MOSAYEBIetal-10} -- lead me to be cautiously optimistic that we now can move systematically toward a theory of glass transitions.  For many years, theorists in this field have felt intuitively that glassy slowing down ought to be accompanied by the growth of some internal length scale. (See \cite{Karmakaretal-13} for a recent review of the search for length scales in glass-forming liquids.)  At the same time, however, it has been the accepted wisdom that glass-forming states remain microscopically liquidlike all the way to their glass transitions.  This wisdom has been based on the observation that density-density correlation functions show no signs of developing long-range order.  Tanaka, in effect, pointed out that one needs to know what to measure.  By looking at correlations between the orientations of different kinds of local topologies, he found evidence that a variety of glass forming systems do indeed exhibit diverging length scales.  Most remarkably, he found that these length scales all seem to be consistent with Ising-like universality.  I describe his results and my proposed interpretation of them \cite{JSL-13} in the later sections of this article.  

Tanaka's results, if confirmed by further investigation, point to the need for  reevaluating much of the existing theoretical work in this field.  To begin that reevaluation, I think we should focus first on the simplest, realistic models of equilibrated, glass-forming liquids, and postpone discussion of the glassy states themselves pending a better understanding of how they are formed. By ``simple'' and ``realistic,'' I mean that these should be models in which classical particles, moving or diffusing at thermally determined velocities, interact with each other via short-ranged forces. Kinetic energies or Brownian diffusion times serve only to set scales of time and pressure for such systems, and thus are essentially irrelevant.  Tanaka's most convincing results are for moderately polydisperse hard disks and spheres, where there is no stored potential energy, and where ordering is controlled only by steric constraints.  This hard-core limit is simpler and more realistic than the mean-field limit that often has been used in this field. 

My physical intuition about disordered systems with short ranged interactions is based on  experiences with the shear-transformation-zone (STZ) theory of nonequilibrium glassy behavior.\cite{FL-11}  This theory accurately accounts for a wide range of rheological phenomena, including  stresses as functions of strain and strain rate, the appearance of yield stresses at low temperatures, shear banding instabilities, oscillatory viscoelasticity, anomalous diffusion, and stretched-exponential relaxation.\cite{FL-11,MANNINGetal-SHEARBANDS-08,BL-11,JSL-12,JSL-EGAMI-12,LIEOU-JSL-12} Several of these developments are directly relevant to glass-forming systems, because they illustrate how broad distributions of time scales emerge in near-equilibrium situations.\cite{BL-11,JSL-12}  A recent application of STZ theory in a nonequilibrium situation is the prediction and experimental confirmation of an annealing-induced, ductile-to-brittle transition in bulk metallic glasses.\cite{RYCROFT-EB-12,SCHROERSetal-BMG-13}  

The STZ theories started from direct observations of bubble rafts, colloidal suspensions, and numerically simulated molecular systems, in which it was clear that the elementary events at the cores of both spontaneous and externally driven fluctuations are local rearrangements of just small numbers of particles.  At very low temperatures, or in otherwise noise-free situations, these rearrangements may occur in spatially extended cascades.  However, at the high densities and intermediate temperatures characteristic of glass-forming systems, they occur at ephemeral flow defects -- the STZ's -- whose population is a fluctuating dynamical feature of any deformable amorphous material.  

Importantly for present purposes, the creation, annihilation, and internal transformations of STZ's are thermally activated, barrier-crossing events; that is, they are transitions between inherent structures.\cite{STILLINGER-WEBER-82,STILLINGER-88}  Trying to compute their rates of occurrence by perturbation expansions in powers of an interaction strength is just as futile as it would be to try to use many-body perturbation theory to predict nucleation rates in supercooled fluids.  Mean-field approximations that are exact only in the limit of infinitely long ranged particle-particle interactions must be used with great caution in these circumstances, if they are to be used at all. Like the Curie-Weiss or Bethe-Peierls approximations, they should be based on well defined, particle-scale physics, and should preserve the length scales that are relevant to the phenomena being described. 

In Sec.~\ref{Theory} of this article, I describe some of the modern glass theories and point out what I perceive to be weaknesses in many of them -- including in one of my own.  The recent simulation results are summarized in Sec.~\ref{Data}.  Being critical of earlier theories in Sec.~\ref{Theory}, I am obliged to present what I hope might develop into a better one.  Accordingly, in Sec.~\ref{Ising}, I summarize my latest proposal \cite{JSL-13} for a glass theory that may be consistent with both the observations and the criteria of simplicity and realism.  I conclude in Sec.~\ref{Conclusions} with summary remarks and a list of questions.

\section{A Critical Review of Glass Theories}
\label{Theory}

It says a great deal about the state of this field that so many imaginative, sensible, but disparate and often contradictory ideas for solving glass problems remain alive today, all with their own advocates.  Each of these ideas contains  elements of the truth. In this key-issues review, I focus primarily on those ideas that seem closest to my main theme, that is, the search for fully thermodynamic theories of glass transitions based on simple but realistic, many-body models.  I therefore give only short shrift to many other important lines of investigation, for which I apologize at the outset. \\

\noindent{\it Kauzmann, Adam, and Gibbs}: The first attempts to make a theory of the glass transition focused on the question of how it might be possible for an ordinary material, obeying classical statistical mechanics, to move toward a frozen state, increasingly unable to access its true equilibrium structure or respond to some kinds of external forcings on observable time scales.  A simple description of this class of phenomena was proposed in 1958 by Adam and Gibbs \cite{ADAM-GIBBS-58}, who wrote the structural relaxation time $\tau_{\alpha}$ in the form:
\begin{equation}
\label{AdamGibbs}
\ln\,\tau_{\alpha} = \ln\,\tau_0 + {{\rm const.}\over T\, s_c(T)},
\end{equation}
where $\tau_0$ is a microscopic time scale, and $s_c(T)$ is the excess entropy per unit volume, measured relative to the entropy of the crystalline state.  Adam and Gibbs interpreted $s_c(T)$ as being inversely proportional to the size of a ``cooperatively rearranging region,'' within which there would be just enough active degrees of freedom, i.e. enough entropy, to enable rearrangements. They further assumed that the excess energy associated with this region scaled with its volume, so that Eq.(\ref{AdamGibbs}) could be interpreted as a thermal activation formula. If, as suggested by Kauzmann, $s_c(T)$ vanishes linearly at $T = T_K$, then Eq.(\ref{AdamGibbs}) becomes the Vogel-Fulcher-Tamann (VFT) relation
\begin{equation}
\label{VFT}
\ln\,\tau_{\alpha} = \ln\,\tau_0 + {T\,D\over T - T_0}
\end{equation}
where $T_0 = T_K$, and $D$ sometimes is known as the inverse fragility. (This definition of ``fragility'' is not exactly the same as the one introduced in \cite{ANGELL-95}, which depends on the definition of $T_g$.)  A number of other forms of these equations have been proposed and tested over the last fifty years.  However, it will be convenient for present purposes to keep Eqs.(\ref{AdamGibbs}) and (\ref{VFT}) as reference points for  the following discussion.\\

\noindent{\it Kinetically Constrained Models}:  The conceptually simplest statistical models that exhibit properties similar to those of Eq.(\ref{VFT}) are the kinetically constrained systems introduced by Fredrickson and Anderson  \cite{FREDRICKSON-ANDERSEN-85} in 1985.  Of these, the most rudimentary example is a two dimensional, non-interacting Ising model in an external field that favors the down spins.  The equilibrium thermodynamics of this model are utterly uninteresting.  However, it exhibits nontrivial glassy dynamics if one adds an artificial kinetic constraint by requiring that a spin can flip, up or down, only if at least two of its neighboring spins are up.  Within that constraint, transition rates can be chosen so that the system approaches a trivial thermal equilibrium.  Numerical simulations of this model reveal a rapidly increasing relaxation time and even an anomalously slow, stretched-exponential decay of fluctuations as the temperature approaches zero.  More sophisticated kinetically constrained models have even more interesting properties; but, in the absence of a direct connection between the kinetic constraints and underlying many-body dynamics, they remain unrealistic.  Recently, however, they have been used in a novel way by Chandler and Garrahan \cite{CHANDLER-GARRAHAN-10}  to study the statistical mechanics of trajectories instead of configurations. It remains to be seen whether this innovative approach will lead to a realistically predictive theory.\\

\noindent{\it Mode Coupling Theory}: The idea that comes closest to being realistic is mode-coupling theory (MCT)\cite{GOTZE-91,GOTZE-92,MC-Cates-09}, which starts with a well posed model of an interacting,   fluidlike, many-body system. MCT is a renormalized, truncated, perturbation theory.  As implied by the remarks in Sec.~\ref{Intro}, it is accurate only so long as the particles are at high enough temperatures and low enough densities that they are weakly scattered while moving past each other.  The theory fails at a mode coupling temperature $T_{MC} > T_0$, where the predicted viscosity diverges. In practice, MCT has been limited by its use of static, two-body correlation functions for information about the many-body interactions.\cite{BERTHIER-TARJUS-11}  More generally, the analytic structure of any perturbation expansion is qualitatively different from that of a theory of activated, barrier-crossing events.  Thus,   there is unlikely to be an accurate way to extend MCT to lower temperatures even by including higher-order correlations. More probably, as happens in other areas of many-body physics, we will have to live with a ``no theory'' region between the mode-coupling and glassy regimes.\\  

\noindent{\it Spin Glasses}: The term ``spin glass'' refers to a class of magnetic alloys in which the interactions between pairs of spins are random and, for the present discussion, can be taken to be equally likely to be ferromagnetic or antiferromagnetic. This class of models was first developed, along with a mean-field solution, by Edwards and Anderson in 1975.\cite{EDWARDS-ANDERSON-75} The simplest spin glass, a cubic Ising model with randomly chosen nearest-neighbor bonds of strength $\pm J$, was shown in a definitive Monte Carlo calculation by Ogielski \cite{OGIELSKI-85} to have a glass transition at $k_B T_0 \cong 1.18\,J$ and to undergo stretched-exponential relaxation at higher temperatures. These models are relevant to the present discussion because they have provided  mathematical examples of how well-posed many-body systems might undergo phase transitions into glasslike states of broken ergodicity.  They are especially important because their behaviors have been used as the starting point for deriving the random first order transition (RFOT) theory of glass transitions. 

In both of the above respects, however, the spin-glass models are manifestly unrealistic.  They are models of systems with quenched disorder -- the distribution over values of the spin-spin interactions is predetermined -- as opposed to being models in which disorder is spontaneously generated, as in glass-forming fluids.  More importantly, our analytic information about the behavior of spin glasses comes largely from mean-field calculations in which the spin-spin interactions are assumed to be infinitely long ranged.  Here, instead of invoking a mean-field approximation for computing averages of local, finite-ranged  interactions, as in \cite{EDWARDS-ANDERSON-75}, it is assumed from the beginning that every spin in the system interacts with every other spin, with coupling strengths chosen at random from the predetermined distribution.  Thus, neither length scales nor even dimensionality play any roles in these theories.  

The spin-glass literature was reviewed comprehensively by Binder and Young in 1986.\cite{BINDER-YOUNG-86}  Much of the theoretical part of their review is devoted to studies of the Sherrington-Kirkpatrick (SK) model \cite{SHERRINGTON-KIRKPATRICK-75}, in which the strengths of infinite-range couplings between Ising spins are chosen from a Gaussian distribution.  The SK solutions, found via both a replica-symmetry-breaking method (Parisi \cite{PARISI-83}) and the  TAP  equations (Thouless, Anderson, and Palmer \cite{TAP-77}), indicate that this model is paramagnetic above a transition temperature, below which it collapses to a state of zero entropy.  

Apart from the entropy collapse, the SK model does not look like a realistic glass.  However, in 1987, Kirkpatrick, Thirumalai and Wolynes \cite{KIRKPATRICK-WOLYNES-87,KIRKPATRICKetal-89}  found a spin-glass model that has more interestingly glasslike properties.  This model is an infinite-range, $p$-state, Potts model, where the two-state Ising spins in the SK model are replaced by $p$-state entities, and $p$ must be at least marginally greater than 4. This model has two phase transitions.  At arbitrarily high temperatures $T$, it is paramagnetic.  When $T$ is decreased to below what is identified, at least formally, as the mode-coupling temperature $T_{MC}$, there appear extensively many stable thermodynamic states, which contribute what is called an excess ``configurational entropy'' to the system.  The number of these states decreases with decreasing $T$ until, at and below a glass transition temperature $T_0$, only a nonextensive number of them remain, and the excess entropy vanishes.  It is this scenario that has been used as a plausibility argument in favor of RFOT. \\

\noindent{\it Random First Order Transition Theory}: At present, RFOT \cite{XIA-WOLYNES-00,LUBCHENKO-WOLYNES-07}   is the most cited of the various glass theories.  It is said to account quantitatively for essentially all glassy phenomena in relatively simple, physically intuitive ways. Thus, it is important to pay serious attention to it.  

In going from the spin-glass results to a glass theory, Wolynes and coworkers assumed that the stable spin-glass states in the temperature range $T_0 < T < T_{MC}$ become metastable  when the interaction range becomes finite, and that they transform among themselves via thermally activated fluctuations.  These authors then hypothesized that the system as a whole consists of a ``mosaic'' of subregions, each characterized by a different one of these metastable states.  In an interpretation suggested by Bouchaud and Biroli \cite{BOUCHAUD-BIROLI-04}, the sizes of these subregions are determined by their stability against the thermally activated fluctuations by which they transform from one state to another. A small subregion will not have enough internal degrees of freedom (in the sense of Adam and Gibbs) to make a transition, whereas one that is too large will break up into smaller ones.  Thus, the mosaic is said to consist of  marginally stable subregions with a $T$ dependent characteristic size, say $R(T)$. The time scale on which these regions transform among themselves is assumed to be the structural relaxation time $\tau_{\alpha}(T)$.  

To estimate $R(T)$ and $\tau_{\alpha}(T)$,  Wolynes and coworkers assumed that the excess free energy of an ``entropically favored droplet'' that might occur in this system can be written in the form
\begin{equation}
\label{RFOT}
\Delta F(R) \approx -\,T\,s_c(T)\,A_d\,R^d + \sigma\,R^b,
\end{equation}
where $s_c(T)$, the excess entropy per unit volume introduced in Eq.(\ref{AdamGibbs}), is assumed to vanish linearly at $T=T_0$.  The quantity $A_d\,R^d$ is the $d$-dimensional volume of a region of linear size $R$.  The second term on the right-hand side of Eq.(\ref{RFOT}) is the energy cost of inserting this droplet into an environment of dissimilar metastable states.  If this were an ordinary surface energy, proportional to a surface area, we would have $b=d-1$. However, throughout the RFOT literature, a variety of rationales have been used to argue that $b = d/2$, which is the value that produces the VFT law.  To see this, note that the resulting $\Delta F(R)$ goes through a maximum at 
\begin{equation}
\label{RFOT1}
R = R^* \sim (T - T_0)^{- 2/d}, 
\end{equation}
and its value at that maximum is 
\begin{equation}
\label{RFOT2}
\Delta F(R^*) \sim (T - T_0)^{- 1}.
\end{equation}
A droplet, i.e. a region, with $R > R^*$ is entropically enabled to grow and, presumably, flow to a different metastable configuration. Thus,  $\Delta F(R^*)$ is the characteristic activation energy for transitions among the metastable states, and the corresponding transition rate is the same as is given in Eq.(\ref{VFT}).  As will be discussed in Sec.~\ref{Data}, the formula for the length scale $R^*$ in Eq.(\ref{RFOT1}) is consistent with Ising scaling.

My most serious question about the RFOT analysis -- more serious than the questions  regarding quenched disorder or $b = d/2$ -- is whether the mosaic picture has any objective reality for glass forming materials with short ranged interactions.  There is a huge, qualitative difference between an infinite-ranged Potts spin glass and a fluid of ordinary particles in an ordinary $d$-dimensional space. Specifically, there is no reason to believe that the statistical physics that produces a multiplicity of thermodynamically stable states in the spin glass is in any way related to the failure of MCT at $T_{MC}$.  As observed earlier, we understand the latter by recognizing that all spontaneous transitions between inherent structures, at low enough temperatures and high enough densities, are thermally activated transitions across energy barriers, and thus are inaccessible via perturbation expansions.  The inherent structures themselves are not thermodynamic metastable states.  On the contrary, true thermodynamic equilibrium is realized as this ergodic system explores the space of inherent structures at rates of the order of the inverse structural relaxation time $\tau_{\alpha}^{-1}$.  

So far as we know experimentally, simple glass-forming materials have at most one metastable state above $T_0$, which is the one that may be metastable against crystallization.  This state can be explored reversibly by varying the system parameters slowly compared to $\tau_{\alpha}^{-1}$. The fact that $\tau_{\alpha}$ increases dramatically with decreasing temperature ultimately must be associated with particle-scale dynamics, which are highly unlikely to have any connection to the structures of infinite-range spin glasses, or even, in my opinion, to the mean-field, density-functional approximations that have been used to derive Eq.(\ref{RFOT}).  Nevertheless, Eqs.(\ref{RFOT1}) and (\ref{RFOT2}), and developments based on them, seem to be consistent with a large amount of experimental data.  We must try to understand why this happens.\\

\noindent{\it Excitation-chain theory}: The XC theory \cite{JSL-XCHAINS-06} was my attempt to find a particle-scale, dynamic mechanism to explain the VFT law.  It was motivated by numerical simulations \cite{GLOTZERetal-XCHAINS-99-00} in which stringy motions, i.e. chains of correlated displacements, seemed to enable transitions between inherent structures.  Like the length $R(T)$ in RFOT, the characteristic size of the chains was predicted to diverge at a transition temperature $T_0$; but my approximation for this behavior did not produce an Ising scaling exponent like that in Eq.(\ref{RFOT1}). So far as I know, this theory is no longer alive; but it illustrated the approach that I had been advocating.\cite{JSL-RefFrame-07}\\    

\noindent{\it Topological Constraints and Jamming}: Both of these theoretical ideas blur the distinction between glass-forming systems and the glassy states themselves, and therefore lie outside the scope of this article.  However, I think that both need ultimately to be incorporated into a larger picture of glass physics, because both involve physically intuitive, diverging length scales.  

There is a large literature on topological constraint theories of molecular glasses, in which the constraints are imposed by chemical bonds.  (See \cite{THORPEetal-2000}, or \cite{MAURO-11} for a concise review.) A central concept in these theories is ``rigidity percolation,'' which is based on the fact that, if the bond lengths are fixed, rigidity of a $d$-dimensional many-body system requires that each molecule be bonded to more than $2 d$ nearest neighbors. Once that requirement is met, a displacement at one point propagates across the system as a whole. Apparently, the best glass formers are those for which this constraint is just marginally satisfied, so that rigidity percolates across large but not infinite distances.  This idea, when elaborated to include bond angles as well as bond lengths, has had impressive practical applications; but it has not yet been pursued by the broader community of glass theorists from a first-principles, statistical point of view. It will be important to find out how to use these topological concepts to develop predictive theories of correlations and relaxation times in glass-forming molecular fluids.

Jamming is a concept that emerges most naturally in theories of granular materials, where temperature is irrelevant, and the structural properties are determined by contact forces between the particles.  In rough analogy to rigidity percolation, the chains of contact forces extend infinitely far in jammed systems, and thus there are naturally diverging length scales as systems approach jamming transitions.  The connections between jamming transitions and glass transitions have been explored by Liu, Nagel, and coworkers, who have proposed interesting scaling relations near what they call ``point J'' in the space of variable density, temperature, and applied stress.\cite{LIU-NAGEL-10}  However, jamming transitions apparently are not exactly the same as glass transitions.  For example, see \cite{IKEDAetal-12}.\\ 

\noindent{\it Stretched-exponential relaxation}:  One of the best known signatures of glassy behavior is the stretched-exponential relaxation (SER) law, according to which the time-dependent decay of many different kinds of perturbations looks like $\exp\,[-\,(t/\tau)^{\beta})]$, instead of like the simple exponential, with $\beta = 1$, expected for linear response functions. Here, the time scale $\tau$ is usually identified as $\tau_{\alpha}$, and the exponent $\beta$ may be substantially less than unity.

This article would be incomplete without mention of Phillips' monumental 1996 review of SER. \cite{PHILLIPS-96}  The main strength of his review is that he looks at an enormous range of experimental observations.  A serious weakness, especially for present purposes, is that he considers only temperatures near or somewhat below $T_g$. His most remarkable result is  that $\beta(T_g)$ is almost always approximately equal to one of  two ``magic numbers,'' $3/5$ and $3/7$.  (See \cite{POTUZAKetal-11} for more recent developments along these lines.)  It seems to me that this work may be undervalued by statistical theorists.  The regularities that Phillips finds in the data might reflect some systematic physics, whether or not they are explained by his theory.  

Phillips' theoretical hypothesis is that glassy relaxation is described by a  diffusion-trap model, in which particles (or some other entities) diffuse in the presence of a random distribution of  absorption centers. The density of particles remaining untrapped after a time $t$ can be shown to decay according to the SER law with $\beta = d^*/(d^*+2)$, where $d^*$ is the dimensionality of the space in which the particles are moving.\cite{GRASSBERGER-PROCACCIA-82} The normal case with $d^* = d = 3$ produces $\beta = 3/5$; a more sophisticated argument involving long-range interactions produces $d^* = 3/2$ and $\beta = 3/7$.  Many questions have been raised about this theory. They start with: What is diffusing?  And, what are the traps?  Since Phillips' analysis pertains only to $T \le T_g$, he can invoke quenched-in heterogeneities to serve as traps.  However, he explicitly declines to use the trap model for $T > T_g$, where the glass-forming fluid presumably is homogeneous, and where we know from experiments and simulations that $\beta$ goes smoothly to unity over a range of values of $T$. He also does not try to use the model to compute the wavenumber dependence of $\beta$ as observed {\it via}  self-intermediate scattering functions, which could be a sharp probe of his underlying physical assumptions.   

The diffusion-trap model is not unique in its ability to describe SER.  In \cite{JSL-12}, I proposed a model of SER that is almost exactly the opposite of the trap model, and which I believe is more realistic.  Instead of assuming that the particles in a glass-forming fluid are free to diffuse until captured by a trap, I assumed that they remain frozen into inherent structures until they are locally rearranged by STZ-like thermal fluctuations.  Using information about STZ's obtained from measurements of oscillatory viscoelasticity\cite{BL-11}, and using a continuous-time random walk analysis, I computed SER curves as functions of both temperature and wavenumber. Perhaps Phillips' magic numbers could emerge from such a calculation.\\

\noindent{\it Other Theoretical Concepts}: The preceding list of theoretical ideas in glass physics is far from complete.  For example, I have not even mentioned the concept of frustration -- the inconsistency between short-range and long-range order that hinders crystallization --  although that concept is implicit in much of the work in this field.  It sometimes has been modeled explicitly as in the work of Kivelson et al.\cite{KIVELSON-95}  Nor have I yet mentioned the increasingly popular term ``dynamic heterogeneity'' \cite{DH-11}; but it appears below in the discussions of correlation lengths in Sec.~\ref{Data} and relaxation rates in Sec.~\ref{Ising}, and I have used it explicitly in discussing SER in \cite{JSL-12}.

\section{Evidence from Numerical Simulations}
\label{Data}

The new information mentioned in Sec.~\ref{Intro} is evidence from numerical simulations reported primarily by Tanaka and coworkers  \cite{TANAKAetal-10,KAWASAKI-TANAKA-11,TANAKA-EPJ-12}, and also by Mosayebi et al.~\cite{MOSAYEBIetal-10}, that {\it may} point to an Ising-like universality in glass forming materials.  I emphasize the uncertainty because the simulations have not yet been confirmed independently, and the results remain controversial. Even if the simulations are exactly correct, they may prove to be relevant only to a small subset of glass forming systems.  Nevertheless, these numerical results are plausible enough to motivate a search for a  theoretical explanation.\\   

\noindent{\it Topological Ordering}: In \cite{TANAKAetal-10,KAWASAKI-TANAKA-11,TANAKA-EPJ-12}, Tanaka et al. report simulations of a variety of two and three dimensional systems, with both hard-core and, in one case, Lennard-Jones interactions. For simplicity, I focus first on their Brownian simulations of moderately polydisperse, hard-core, colloidal suspensions, where temperature is irrelevant, and where the approach to the glass transition is controlled only by the volume fraction (density) $\phi$. They looked for spatial correlations, not between particle positions {\it per se}, but between the positions of particles in topologically similar environments. Specifically, for two-dimensional hard-disks, they measured time-averaged, hexatic order parameters $\bar\psi_6$ as functions of position, and computed two-point correlations $\langle\bar\psi_6(r)\,\bar\psi^*_6(0)\rangle$ as functions of the separation $r$.  From the latter quantity, they deduced a correlation length $\xi(\phi)$, and found by a finite-size scaling analysis that it was proportional to $t^{-\nu}$, where $t \equiv (\phi_c - \phi)/\phi$, and the critical volume fraction $\phi_c$ depends on the degree of polydispersity $\Delta$ (the percentage width of a Gaussian distribution over particle radii).  Similar results were obtained for polydisperse hard spheres in three dimensions, where the relevant topological order parameter was found to be the degree of hexagonal-close-packed  (as opposed to icosahedral) order. (See also \cite{LEOCMACH-TANAKA-12}.) In both cases, they found that $\nu \cong 2/d$, which is indistinguishable from the Ising hyperscaling relation $\nu = 2/d - \alpha$, because the specific heat exponent $\alpha$ is negligibly small for these purposes. 

Tanaka and coworkers  \cite{TANAKAetal-10,KAWASAKI-TANAKA-11,TANAKA-EPJ-12} also have measured structural relaxation times $\tau_{\alpha}$ for each of the various systems that they studied. For both $d = 2 ~{\rm and}~ 3$, their results are consistent with the Vogel-Fulcher-Tamann (VFT) relation, 
\begin{equation}
\label{VFTxi}
\ln\,(\tau_{\alpha})\sim \xi^{d/2} \sim t^{-1}, 
\end{equation}
where $\xi$ is their measured correlation length. As emphasized by Tanaka in \cite{TANAKA-EPJ-12}, the dramatic slowing down near glass transitions described by Eq.~(\ref{VFTxi}) has prevented simulations from coming sufficiently close to critical points to confirm the apparent limiting behaviors. The growth of correlations has been confirmed out to only one decade at best. However, the consistency of these results over the range of different models and system parameters, and the apparent Ising-like universality, makes it hard to resist taking this data seriously, at least pending further study.\\ 

\noindent{\it Dynamic Correlations}: As part of their series of investigations, Kawasaki and Tanaka \cite{KAWASAKI-TANAKA-11} measured what is known as the ``dynamic correlation length'' $\xi_D$, which plays an important role in the following discussion.  This length scale has emerged in analyses of dynamic heterogeneities, i.e. the spatial inhomogeneities that are observed in the dynamic behaviors of glass-forming systems. \cite{DH-11}  It is defined most simply as follows.  Choose two points separated by a distance $r$.  Compute the probability that neither of the particles near those two points has moved out of its local environment (its "cage") after a time of the order of the structural relaxation time $\tau_{\alpha}$.  That probability decays as a function of increasing $r$, apparently like $\exp( - r/\xi_D)$.  Thus, $\xi_D$ is a rigidity length, roughly analogous to the length discussed in topological constraint theories. It often is claimed in the literature (see below) that this dynamic length scale need not have a structural origin. I find that assertion hard to believe. 

In  \cite{KAWASAKI-TANAKA-11}, Kawasaki and Tanaka report a parallel study of polydisperse and bidisperse hard-core colloids in two dimensions.  They find that their bidisperse systems, unlike the polydisperse ones, do not exhibit long-range hexatic correlations.  However, their measured values of $\xi_D$ do seem to diverge with the Ising-like exponent, and to be consistent with the Vogel-Fulcher-Tamann formula in Eq.~(\ref{VFTxi}). \\ 

\noindent{\it Non-Affine Displacements}: An independent analysis by Mosayebi et al.~\cite{MOSAYEBIetal-10}, for a three dimensional, bidisperse, Lennard-Jones system, adds weight to the evidence for Ising-like behavior.  Here, Ising-like correlations (with $\nu \cong 2/3$) were observed in the  non-affine parts of the displacements induced by small, applied strains. These authors also approximately confirmed the VFT formula for $\tau_{\alpha}$ for their model.  Their results, in combination with those presented in  \cite{KAWASAKI-TANAKA-11}, imply that the structural ordering mechanisms in bidisperse systems are qualitatively different from those in moderately polydisperse systems. \\ 

\noindent{\it Equations of State}: Yet more evidence bearing on the phase transitions that occur in polydisperse hard-disk systems is contained in \cite{KAWASAKI-TANAKA-11}, where the authors report measurements of the equations of state, i.e. the pressures $p$ as functions of $\phi$, for  a sequence of increasing percentage polydispersities $\Delta$. As expected, the monodisperse system at $\Delta = 0\,\%$ exhibits a transition between liquid and hexatic phases at $\phi \cong 0.69$.\cite{JASTER-99,ANDERSONetal-12}  With increasing $\Delta$, the transition points on the $p(\phi)$ curves move to larger $p$'s and $\phi$'s, and become less and less distinct.  They are invisible in the pressure data above $\Delta = 9 \%$, which is the value of the polydispersity for which  Tanaka et al.~\cite{TANAKAetal-10} report an Ising-like bond-orientational correlation length that extrapolates to infinity at $\phi = \phi_c \cong 0.787$, and a corresponding divergence of $\tau_{\alpha}$. At larger values of $\Delta$, they see evidence in the form of diverging relaxation times for glass transitions at larger values of $p$ and $\phi$. The important point for present purposes is that the sequence of pressure curves described in \cite{KAWASAKI-TANAKA-11} indicates a smooth crossover from a liquid-hexatic transition at $\Delta = 0\,\%$ to Ising-like critical points for $\Delta \ge 9\,\%$ -- a qualitative change of universality class.\\  

\noindent{\it Point-to-Set Calculations}: There is another body of numerical data that appears to be  inconsistent with the diverging correlation lengths reported in  \cite{TANAKAetal-10,MOSAYEBIetal-10}.  I refer here to ``point-to-set'' calculations, which started as attempts to find direct evidence for the RFOT mosaic structure based on a mathematical construction proposed by Bouchaud and Biroli.\cite{BOUCHAUD-BIROLI-04}  The idea is to compare a freely equilibrated configuration with a corresponding configuration that is equilibrated after the positions of some set of the particles have been fixed.  Supposedly, the difference between the configurations should disappear as the distance between the observation point and the positions of the fixed set of particles becomes larger than any length scale in the system. The hope is that this procedure yields ``order-agnostic'' many-body information beyond that given by positional pair correlations, and that this information can be used to deduce the length scale associated with the mosaic pattern as interpreted in \cite{BOUCHAUD-BIROLI-04}.  

Recent examples of such point-to-set calculations include those of Berthier and Kob \cite{BERTHIER-KOB-12}, and of Charbonneau and Tarjus \cite{CHARBONNEAU-TARJUS-13}, both of which use binary mixtures of spherical particles.  Both produce only weakly growing static length scales that are smaller than the dynamic lengths $\xi_D$ that were measured for the same systems. These authors conclude that the static and dynamic behaviors of these glass-forming systems must somehow be decoupled from each other. If this is true, then the diverging static correlations found in  \cite{TANAKAetal-10} for polydisperse systems can be, at best, properties of only a very special, non-characteristic class of models; and the results reported in  \cite{MOSAYEBIetal-10} must somehow be wrong. 

Charbonneau and Tarjus \cite{CHARBONNEAU-TARJUS-13} pay special attention to an inequality derived {\it via} a lengthy point-to-set analysis by Montanari and Semerjian \cite{MONTANARI-SEMERJIAN-06}, which states that 
\begin{equation}
\label{MSinequality}
\tau_{\alpha} \le \tau_0 \,\exp\, ({\rm const.}\times\,\xi^d),
\end{equation}
where $\xi$ is supposedly the same static length that is determined by the point-to-set method.  However, as argued in the introductory paragraphs of  \cite{MONTANARI-SEMERJIAN-06}, the MS inequality in Eq.(\ref{MSinequality}) is a general relation between $\tau_{\alpha}$ and whatever $\xi$ is the longest correlation length in the system.  To see this, think of the system as consisting of independently fluctuating regions of size $\xi$, and note that the slowest possible relaxation mechanism in any such region would be a thermally activated process whose largest possible activation energy (for particles with finite-ranged interactions) would scale like $\xi^d$.  With this interpretation, Eq.(\ref{MSinequality}) tells us that a diverging time scale requires a diverging length scale of some kind.  

The MS inequality seems hard to reconcile with the point-to-set results reported in  \cite{CHARBONNEAU-TARJUS-13}, where $\tau_{\alpha}$ grows too rapidly to look consistent with the measured values of $\xi$, at least within the range of validity of the simulations.  Charbonneau and Tarjus tentatively attribute this inconsistency to unobservable complexities in the approach to the glass transition.  I suspect, however, that there is a simpler explanation.  

The numerical evidence presented in  \cite{BERTHIER-KOB-12,CHARBONNEAU-TARJUS-13}, combined with the MS inequality and the fact that the measured $\xi_D$ is consistent with Eq.(\ref{VFTxi}), leads me to suspect that the position-based point-to-set calculations are not revealing much more than the well known absence of long-ranged density-density pair correlations.  For the sake of argument, consider Tanaka's system of polydisperse hard disks, and suppose that the particle positions used in a point-to-set analysis are replaced as local state variables by the hexatic order parameters $\psi_6(r)$. We could fix the values of some subset of them, and then ask how their values at other points depend on the fixed set.  The resulting length scale would almost certainly be the same as the diverging correlation length found by Tanaka et al., because the point-to-set result would be dominated by the hexatic pair correlations; and the MS inequality would easily be satisfied. 

Then note that, because $\tau_{\alpha}$ diverges for the binary mixtures studied in  \cite{BERTHIER-KOB-12,CHARBONNEAU-TARJUS-13}, the MS inequality implies that there must be some diverging structural correlation length in those systems, perhaps the same as the one seen in \cite{MOSAYEBIetal-10}, and perhaps equal to $\xi_D$; and thus there must be some analogs of $\psi_6(r)$ that we have yet to discover.  If this line of reasoning is correct, it obviates any need for the point-to-set analysis or, for that matter, any need for the RFOT mosaic hypothesis.

\section{Ising-Like Description of Glass Forming Fluids}
\label{Ising} 

I turn now to a theoretical interpretation of the observations of Tanaka et al.~\cite{TANAKAetal-10,KAWASAKI-TANAKA-11}  This theory is presented fully in \cite{JSL-13}; what follows is a summary that focuses on the physical ideas rather than technical details.  For clarity, I describe the theory in terms specific to Tanaka's two dimensional, polydisperse, hard-disk model, so that I can relate the ideas to the direct image of that model shown here in Fig.~\ref{ROP1}.  I emphasize, however, that the mathematics and the concepts are equally valid in three dimensions and for other topological orderings.  As presently constructed, the theory is intrinsically  universal. 

\subsection{Two-state clusters}

\begin{figure}
\centering \epsfig{width=.5\textwidth,file=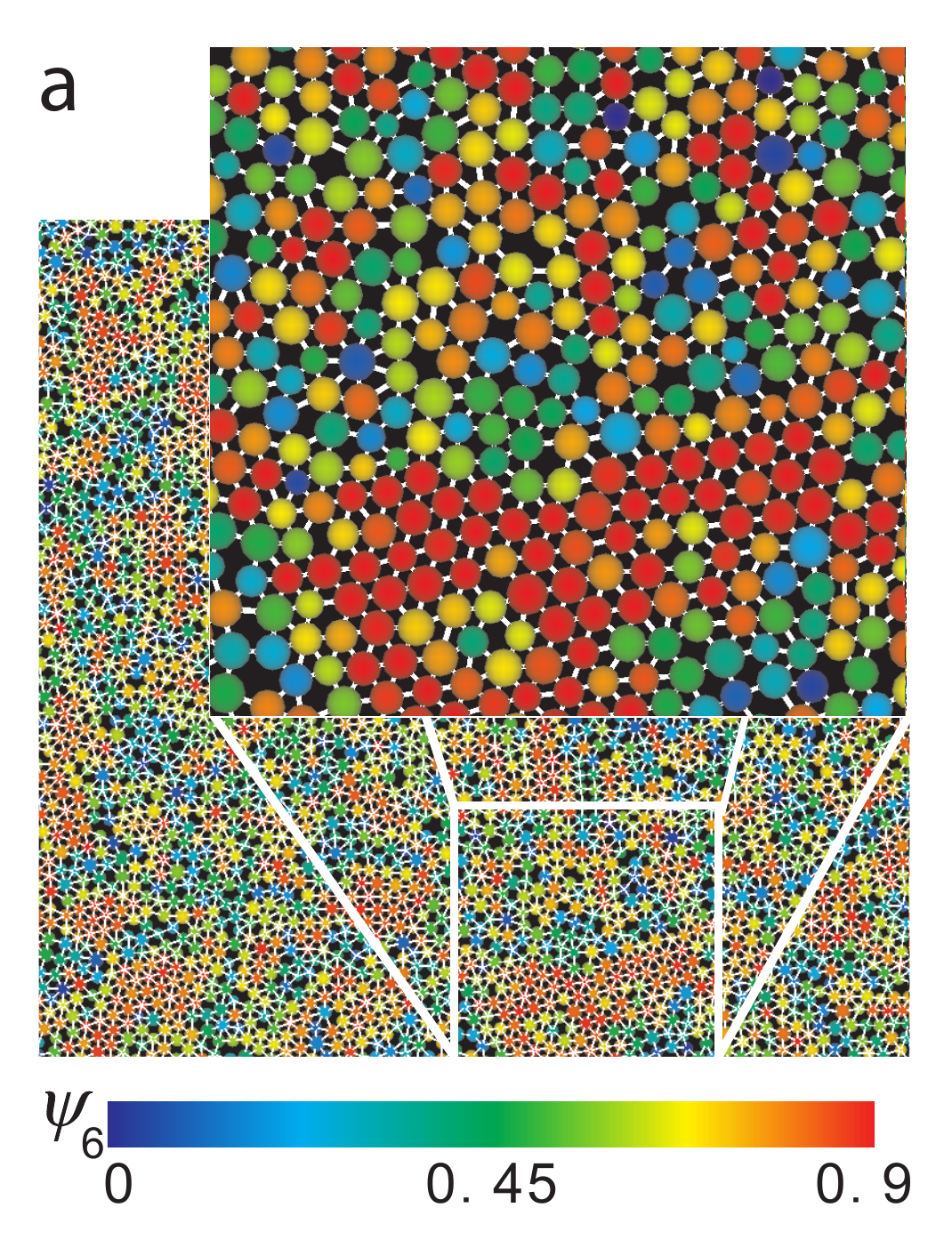} \caption{Figure 2a from Tanaka et al. \cite{TANAKAetal-10}.  This is an instantaneous configuration of a system of hard disks with polydispersity $\Delta = 9 \%$, at volume fraction $\phi = 0.73$.  The degree of local hexatic order $\psi_6$ for each particle is indicated by the color scale, red (darker) for large $\psi_6$ and green (lighter) for small $\psi_6$. Nearest-neighbor ``bonds'' are shown by thin white lines.} \label{ROP1}
\end{figure}

Figure 2a in \cite{TANAKAetal-10}, reproduced here as Fig.~\ref{ROP1}, is a snapshot of an instantaneous configuration of a system of hard disks with polydispersity $\Delta = 9 \%$, at volume (i.e. area) fraction $\phi = 0.73$.  The upper, right-hand part of the figure is a magnification of the square outlined by the white lines at the bottom.  The color scale indicates the degree of local hexatic order $\psi_6$ for each particle. The short white lines are nearest-neighbor ``bonds'' deduced from a Voronoi construction. 

%%%%%%%%%%%%% FIGURE 1 %%%%%%%%%%%%
%%%%%%%%%%%%%%%%%%%%%%%%%%%%%%%%%%%%% 

The most obvious feature of this picture is the pattern of hexatically ordered (red) regions in the midst of less hexatic (green) particles and (black) voids.  The red regions are not growing crystallites.  On the contrary, they are internally correlated areas, whose sizes are of the order of the correlation length $\xi$, that are appearing and disappearing during the normal fluctuations of a thermodynamically equilibrated state. Close inspection of the figure reveals that separated red regions are not aligned with each other.  The maximum misalignment for hexagons is only $\pm 30^{\circ}$; thus, this feature of the figure is not immediately obvious.  It is important, however, because it tells us that the orientational correlations do not extend beyond the individual red regions.  

In the absence of interaction energies, the only relevant, extensive, thermodynamic variable for this system is its volume $V$.  (For convenience, I use the term ``volume'' to mean either three dimensional volume or, as in the present case, two dimensional area.)  In  ordinary, energetically controlled systems, our intuition tells us that stable states occur at minima of the energy or, more precisely, of the free energy. As the temperature decreases, low energy configurations become increasingly probable.  The analogy here, where only steric constraints are operative, is that the stable configurations are most probably those with the smallest volumes.  Thus, relatively compact hexatic order is favored for hard disks, which is partly what we are seeing in Fig.~1.  Moreover, as the pressure and density increase, these more compact configurations become inceasingly prevalent. 

To make the preceding discussion more specific, note that Fig.~\ref{ROP1} can be interpreted as an array of two distinct kinds of small clusters: those that have hexatic topologies and those that do not.  An hexatic cluster consists, minimally, of a central particle and  six, almost regularly spaced, nearest neighbors. The other clusters are less regular and may, or may not, contain voids. The hexatic clusters are more compact than the non-hexatic ones; thus, we expect the hexatic regions to grow with increasing $p$ and $\phi$.  But this is only a part of the story; it does not explain the orientational correlations between the hexatic clusters. 

The correlation theory presented in  \cite{JSL-13} is based on two central arguments.  First, the hexatic clusters necessarily have spatial orientations, and the volume that they and their nearest neighbors occupy is smaller if neighboring clusters are aligned with each other than if they are misaligned.  In other words, the volume is minimized when the hexatic clusters fit together in an orientationally aligned array.  Second, and most crucially, the hexatic clusters are two-state systems. The first of these arguments seems trivially obvious, because misalignment destroys the hexatic order in the space between the clusters. The second argument is central to the Ising analogy, and needs careful consideration. 

On the average, a system of polydisperse hard disks is rotationally symmetric; but any single hexatic cluster, sitting at a given position in a disordered environment, does not see rotational symmetry.  In order to participate in an ordering transition, however, this cluster must have some orientational flexibility.  It must be able to realign itself in the presence of other oriented clusters, which means that it must have at least two orientations in which it is almost equally comfortable.  In a dense, disordered environment, the probability of there being more than two such  favorable orientations is negligibly small; thus, the statistically relevant hexatic clusters are two-state systems.  The two-state idea goes back at least as far as the 1972 paper by Anderson, Halperin and Varma.\cite{AHV-72} Note the essential role played by disorder in this argument.  

To describe this picture mathematically, let $N_+$ and $N_-$ be extensive, internal variables denoting the numbers of hexatic clusters oriented in $+$ and $-$ directions with respect to some direction in space; and let $N_0$ denote the number of non-hexatic clusters, which, in  \cite{JSL-13}, are simply called ``voids.'' As in the STZ theory, the actual orientations denoted by $\pm$ need not be specified initially.  The various orientations of the ordered regions seen in Fig.~\ref{ROP1} are local, spontaneously broken symmetries; and $\pm$ can be understood as referring to those local orientations. 

In terms of these internal variables, the volume of the system, up to an additive, $\Delta$-dependent constant, is 
\begin{equation}
\label{V}
{\cal V} \cong N^*\,v^* + N_0\,v_0 - {J\over 2\,(N^* + N_0)}\,(N_+^2+N_-^2),
\end{equation}
where  $N^* =N_++N_-$ is the total number of hexatic clusters, $N_{\pm}/(N^* + N_0)$ is the density of $\pm$ clusters, $v^*$ is the volume of an hexatic cluster (of either orientation), and the term proportional to $J$ is the conventional mean-field approximation for the near-neighbor interaction between these clusters.  By construction, $J$ must be a non-negative constant.  The quantity $v_0$ is the volume of a void. To see the analogy between Eq.(\ref{V}) and an Ising system, define 
\begin{equation}
\label{m-eta-def}
m = {N_+ - N_-\over N^*},~~~ \eta = {N^*\over N^* + N_0}.
\end{equation}
The ``magnetization'' $m$ is the bond-orientational order parameter. $\eta$ measures how close the system is to its maximum density; it vanishes in the dilute limit, $N_0 \to \infty$, and goes to unity at high density where the voids are squeezed out of the system. With this change of variables, Eq.(\ref{V}) becomes
\begin{equation}
\label{calV}
{{\cal V}(m,\eta)\over N^*}=  v^*  +\left({1\over \eta} - 1\right)\,v_0 - {1\over 4}\,J\,\eta\,(1 + m^2).
\end{equation}
Note that the term proportional to $J$ contains a factor $\eta$, implying that ordering becomes weaker with increased numbers of voids. 

The easiest way to use Eq.(\ref{calV}) in a glass theory is to look for states of maximum entropy  subject to the condition of fixed  volume ${\cal V}(m,\eta)$. To do this, look for minima of the ``free volume''
\begin{equation}
\label{calF}
{\cal F}(m,\eta) = {\cal V}(m,\eta) - X\,{\cal S}(m,\eta),
\end{equation}
where $X$ is a Lagrange multiplier, and ${\cal S}(m,\eta)$ is the dimensionless entropy (the logarithm of some measure of the number of states). In the absence of potential energy, the ordinary free energy is simply $-\,k_B T\,{\cal S}$; thus, differentiating Eq.(\ref{calF}) with respect to ${\cal V}$, we find that $1/X = \partial {\cal S}/\partial {\cal V} = p/k_B T$.  

Next, in the spirit of the Ising analogy, assume that the $m$-dependence of the entropy ${\cal S}(m,\eta)$ can be computed, as usual, by counting the ways in which $N_+$ ``up'' states can be distributed among $N^* =N_++N_-$ sites.  Then, minimize the resulting ${\cal F}(m,\eta)$ with respect to $m$.  The result looks like -- and, indeed, is -- a Curie-Weiss mean-field formula:
\begin{equation}
\label{mequation}
m = \tanh\left({p\,J\,\eta\,m\over 2\,k_B T}\right).
\end{equation}  
The most important implication of Eq.(\ref{mequation}) is that this model has a mean-field critical point at $\eta_c = 2\,k_B T/p\,J$, such that the Ising symmetry under $m \to -\,m$ is spontaneously broken for $\eta > \eta_c$, and where $m$ undergoes critical fluctuations. It is easy to check that the fluctuations in $\eta$ remain non-critical.    

In \cite{JSL-13}, I assumed an explicit form for the $\eta$-dependence of ${\cal S}(m,\eta)$ and computed minima of ${\cal F}(m,\eta)$ in the space of variables $m$ and $\eta$.  By doing this, and invoking a renormalization-group analysis, I confirmed that this procedure recovers the equations of state, $p(\phi)$, reported by Kawasaki and Tanaka \cite{KAWASAKI-TANAKA-11} and described here in Sec.~\ref{Data}.  This analysis served as a quantitative self-consistency check on the theory. It also confirmed that $J$ is a decreasing function of increasing $\Delta$, in accord with the idea that the ordering strength decreases with increasing polydispersity.

To complete the calculation of the correlation length $\xi$, at least in principle, we should let $m$ be a function of position, and generalize ${\cal F}(m,\eta)$ to include a square-gradient term proportional to $J\,({\bf \nabla} m)^2$.  We then should use the functional $\exp\,(- {\cal F}/X)$ as a statistical weight in a function space, and perform a renormalization-group calculation to obtain equations of state and correlation functions.  (For example, see \cite{GOLDENFELD-92}.) But we know, just from the Ising symmetry of this theory, that the correlations computed in this way will be those described here in Sec.\ref{Data}.  

\subsection{Relaxation rates}

If the diverging glassy length scale $\xi$ is an equilibrium property, as opposed to an intrinsically dynamic one, then we still need to understand how it determines a diverging time scale.  The following discussion, like the preceding one, is taken from  \cite{JSL-13}.  

Note first that, although the equilibrium glass transition appears to occur at an Ising-like critical point with an Ising-like divergence of spatial correlations, the dynamic critical behavior of the glass is qualitatively different from that of an Ising magnet. The scaling analysis by Hohenberg and Halperin \cite{HOHENBERG-HALPERIN-77}  tells us that the relaxation time for fluctuations of a non-conserved Ising magnetization diverges relatively weakly, like a power of $\xi$; whereas the VFT law tells us that $\tau_{\alpha} \sim \exp(\xi^{d/2})$.  The difference is that relaxation events in a glass-forming fluid near its transition point are highly nonlinear collective phenomena, not amenable to the perturbation-theoretic methods or the assumptions about the nature of noise sources implicit in  \cite{HOHENBERG-HALPERIN-77} or in mode coupling theory.\cite{GOTZE-91,GOTZE-92}

As before, assume that structural rearrangements in glass forming materials occur at STZ's or at other similarly soft, local fluctuations.  If the characteristic formation volume of an STZ is $v_Z$ (roughly a single-particle volume), then the equilibrium STZ population is proportional to a Boltzmann factor $\exp\,(-\,v_Z/X)$, where $X = k_B T/p$.  To estimate a spontaneous STZ formation rate, and thus a relaxation rate, multiply this Boltzmann factor by an attempt frequency, $\rho(X)/\tau_0$, where $\tau_0$ is a microscopic time determined by the kinetic energies of the particles or the thermal fluctuations of the fluid in which they are suspended. The dimensionless attempt frequency $\rho(X)$ describes glassy slowing down as $X$ decreases, i.e. as $p$ increases. It is proportional to $\tau_0/\tau_{\alpha}$; its evaluation is the goal of any glass theory.

Kawasaki and Tanaka  \cite{KAWASAKI-TANAKA-11} show by direct imaging that relaxation events occur primarily in disordered regions, consistent with the observation of Widmer-Cooper and Harrowell \cite{HARROWELL-07} that particles undergo rearrangements in regions of high ``propensity.'' In the present picture, this observation means simply that the STZ formation volume $v_Z$ is smaller in the more loosely connected disordered regions than in the ordered ones, so that the STZ's appear most frequently in the former.  However, the attempt frequency $\rho(X)$ must involve collective motions of large numbers of particles, rather than being determined by the local environments of just a few of them. 

The correlated regions of size $\xi$ shown in Fig.~\ref{ROP1} are slowly fluctuating into and out of existence at a rate that I identify as being proportional to $\rho(X)/\tau_0$.  The STZ transitions provide the mechanisms by which these fluctuations occur; conversely, it is these collective fluctuations that self-consistently create and annihilate the STZ's. To estimate this rate, note that a correlated volume ${\cal V}_{corr}$ of linear size $\xi$ contains a number of particles proportional to $\xi^d$. In a thermally fluctuating system, each of these particles makes small, independent, forward and backward displacements through distances of the order of the interparticle spacing. Therefore,  ${\cal V}_{corr}$ undergoes Gaussian fluctuations of a characteristic magnitude $\delta\,{\cal V}_{corr}$ proportional to the square root of its size; that is, $\delta\,{\cal V}_{corr} \sim \xi^{d/2}$. To estimate a time scale for these fluctuations, note that they are slow, activated events.  Therefore, the statistical analysis in \cite{LIEOU-JSL-12,BLI-09} tells us that their frequency is proportional to $\rho(X)$, where
\begin{equation}
\label{rho-xi}
-\,\ln \rho(X) \sim  {\delta{\cal V}_{c\!o\!r\!r}\over X} \sim {\xi^{d/2}\over X_c} \sim {1\over t^w}, 
\end{equation}
where $X_c$ is the critical value of $X$ and $w = d\,\nu/2= 1- \alpha/2 \cong 1$ for both $d= 2$ and $3$. Thus, we recover the VFT formula.  

We can push the argument leading to Eq.(\ref{rho-xi}) a bit further by noting that it implies
\begin{equation}
\label{VFT2}
\ln\left({\tau_{\alpha}\over \tau_0}\right) \approx {D\,\phi\over \phi_c - \phi};~~~ D = p_c\,\xi_0^{d}/k_B T,
\end{equation} 
where $\xi_0$ is a length proportional to the particle spacing, and $p_c$ is the critical pressure.   We know from \cite{KAWASAKI-TANAKA-11}  that $p_c$ increases with $\Delta$.  Thus, the inverse fragility parameter $D$ is predicted to increase with $\Delta$ -- the glass becomes stronger -- in agreement with the simulations.

\section{Concluding Remarks}
\label{Conclusions}

My main theme in this paper is that, to make progress in understanding the glassy state of matter, we should pay closer attention to simple, realistic models of thermally equilibrated, glass-forming materials. Tanaka and his coworkers have shown us some nontrivial examples in which favored local topologies in disordered fluids collectively produce long-range correlations that apparently extrapolate to Ising-like critical points.  I have suggested a general mechanism by which a disordered environment might cause these local topologies to align with each other as if they were, effectively, two-state entities, consistent with Ising universality.  

So far, these pieces of the puzzle seem to be fitting together. The analysis leading to Eq.~(\ref{VFT2}) recovers the central RFOT results summarized by Eqs.~(\ref{RFOT1}) and (\ref{RFOT2}).  Unlike RFOT, the Ising analysis starts by finding an observable equilibrium correlation length $\xi$, rather than by deducing a length scale $R$ from a hypothetical dynamic mechanism. Moreover, Tanaka's picture of fluctuating regions of bond-orientational order, seen in Fig.\ref{ROP1}, looks qualitatively different from the fluctuating mosaic structure postulated by RFOT \cite{LUBCHENKO-WOLYNES-07} and interpreted by Bouchaud and Biroli.\cite{BOUCHAUD-BIROLI-04}.  

All of the Ising ideas need to be carefully questioned, especially where the basic assumptions differ from those of other theories. I close by listing some of the questions that I consider to be most urgent.

How {\it generally} valid is Tanaka's picture of correlations?  How general is his idea of bond-orientational order?  If bidisperse systems \cite{MOSAYEBIetal-10} exhibit Ising-like correlations but no bond-orientational order \cite{TANAKA-EPJ-12}, what kind of structural ordering might be occurring?  What other possibilities are there?  

Are the point-to-set calculations truly order agnostic?  Does their failure to reveal long-range structural correlations in bidisperse systems really mean that no such correlations exist, despite the evidence for long-range dynamic correlations?  Conversely, are the point-to-set methods powerful enough to detect long-range topological order in moderately polydisperse systems, where it is almost certain that such correlations do exist?  

Might the dynamic length $\xi_D$ be a truly order-agnostic indication of structural correlations? When $\xi_D$ appears to diverge at a glass transition, must there always be a correspondingly divergent structural correlation that can be understood directly in terms of physical many-body interactions?  Or might dynamic correlations somehow be decoupled from structural (i.e. static, equal-time) correlations?

What about molecular glasses with finite-ranged interaction potentials?  Or network glasses with true chemical bonds? It might seem easy to translate the volume-based statistical analysis described here into a more conventional energy-based theory.  Is it really easy?  How might rigidity percolation appear in such a theory?  

How {\it precisely} valid are Tanaka's results, if only for the polydisperse hard-core systems near their critical points?  We know that, for weak or vanishing polydispersity, the two-dimensional hard-disk system undergoes a liquid-to-hexatic transition at which the correlation length diverges  more abruptly than it does for an Ising system.  (This is a ``Kosterlitz-Thouless-Halperin-Nelson-Young'' two-dimensional melting transition.\cite{KOSTERLITZ-THOULESS-72,KOSTERLITZ-74,HALPERIN-NELSON-78,NELSON-HALPERIN-79,YOUNG-79} See \cite{JSL-13} for a discussion of the crossover between it and the glass transition.)  Might something like this happen at larger polydispersities, in a way that would not be detectable by Tanaka's finite-size scaling analyses?  What happens at even larger polydispersities, where the systems become so strongly disordered that the coupling between oriented clusters, if such clusters exist at all, becomes vanishingly weak?  Are these glasses?  Or something else?  

How literally can we take Tanaka's fits to the VFT formula near his critical points? Or my derivation of it in Eq.(\ref{rho-xi})?  The VFT formula has not always been a reliable approximation for $\tau_{\alpha}$ in comparisons with  experimental or numerical data. For example, in recent applications of STZ theory to nonequilibrium situations \cite{LIEOU-JSL-12,JSL-EGAMI-12}, my coworkers and I have found it better to determine $\rho(X)$ from the data than to try to predict it from the VFT formula.  On the other hand, there are serious theoretical uncertainties in trying to deduce values of $\tau_{\alpha}$ from measurements, say, of viscosity or diffusion.\cite{JSL-12}  

Until now, I have followed my own advice and have avoided talking about the glassy states that are formed below the glass temperature or above the analogous volume fraction; but, eventually, we will need to pay attention to them.  These are nonequilibrium states whose properties must depend on their histories of formation. According to the Ising theory, they must include quenched-in  heterogeneities such as boundaries between regions of different partial orderings, or clusters of voids.  More generally, their states of internal disorder are determined by quench rates, aging times, and the like.  A growing understanding of the glass-forming states should guide us in predicting the properties of the glasses themselves.  This is already happening, for example, in the recent predictions of the fracture toughness of annealed bulk metallic glasses.\cite{RYCROFT-EB-12}  

In short, we need to understand the extent to which the hard-core, Ising-like system discussed here can -- or cannot -- serve as a paradigm for understanding the much larger world of realistic glassy materials.

\begin{acknowledgments}

I thank H. Tanaka for providing Fig.~\ref{ROP1} and for highly informative discussions. This research  was supported in part by the U.S. Department of Energy, Office of Basic Energy Sciences, Materials Science and Engineering Division, DE-AC05-00OR-22725, through a subcontract from Oak Ridge National Laboratory.

\end{acknowledgments}

\end{document}